\documentclass{elsart}
\usepackage{graphicx}
\usepackage{amssymb}

\begin{document}
\begin{frontmatter}

\title{Structural, Vibrational and Mechanical Studies of Hydroxyapatite produced by 
wet-chemical methods}
\author{K. Donadel},
\author{M. C. M. Laranjeira\corauthref{cor1}},
\ead{mauro@qmc.ufsc.br}
\corauth[cor1]{Corresponding author.}
\author{V. L. Gon\c{c}alves},
\author{V. T. F\'avere},
\address{Depto de Qu\'{\i}mica, Universidade Federal de Santa Catarina, Florian\'opolis, 
Santa Catarina, Brazil, Cx. P. 476, 88040-900}
\author{K. D. Machado},
\ead{kleber@fisica.ufsc.br}
\author{J. C. de Lima},
\address{Depto de F\'{\i}sica, Universidade Federal de Santa Catarina, Florian\'opolis, 
Santa Catarina, Brazil, Cx. P. 476, 88040-900}
\author{L. H. M. Prates}
\address{Depto de Estomatologia, Universidade Federal de Santa Catarina, Florian\'opolis, 
Santa Catarina, Brazil, Cx. P. 476, 88040-900}

\begin{abstract}
Hydroxyapatite samples were produced by two different wet-chemical methods, and 
characterized by x-ray diffraction, infrared and compression strength measurements. The 
x-ray diffraction measurements were simulated using the Rietveld method, and structural 
data as lattice parameters and average crystallite size were obtained. The infrared 
spectra showed the presence of CO$_3^{2-}$ ions in all samples, indicating a 
contamination by these ions. By mixing samples produced by both methods, a bioceramic  
was obtained and, after sintering, samples with very high compression strengths (26--30 MPa) 
were obtained.
\end{abstract}

\end{frontmatter}

\section{Introduction}

Calcium hydroxyapatite, Ca$_{10}$(PO$_4$)$_6$(OH)$_2$ (CaHAp), is the main inorganic 
component of hard tissues of the bones of vertebrates. It answers for 60--70\% of the 
mineral phase in the human bone \cite{Gil,Bayraktar,Finisie}, and 
it is a member of the `apatite' family of 
compounds. Biological apatites, which comprise the mineral phases of tissues like enamel, 
dentin and bone, usually have compositions and crystallinity slightly different from the 
stoichiometric CaHAp. These structural differences are responsible for different physical 
and mechanical properties found for these materials. In the body, they were usually 
observed to be carbonate (CO$_3^{2-}$) substituted and calcium deficient \cite{Hench}. 
Bone tissue 
has the ability to regenerate, forming healthy tissue that grows in the direction of 
the damaged tissue in order to repair it. So, if a material with physical and chemical 
properties similar to the biological bone constituents could be produced, the growth 
process of healthy tissue can be accelerated and the recovery time of the patients can 
be decreased, for instance. Such materials should also have porosity and compression 
strength similar to the tissues they will substitute, in order to assure 
biocompatibility. Successful experiments on animals (rats, rabbits and dogs) led to 
clinical application of CaHAp bioceramics in humans over the last two decades. These 
experiments included the grafting of periodontal defects, post-traumatic long bone 
defects and augmentation of the alveolar ridge and maxillofacial skeleton 
\cite{Rejda,Jarcho,Rosen,Passuti,Byrd,Satoh}.

To produce synthetic CaHAp with the desired properties, chemical and hydrothermal 
methods can be used \cite{Gil,Tadic,Morales}. In general, these methods allow the production 
of materials 
with good crystallinity, physiological stability and with the morphological 
characteristics of the hard tissues, but some of the physical, chemical and mechanical 
properties of the final products usually depend on the specific method used in the 
synthesis. In addition, contaminant phases, such as calcium oxide (CaO) and tricalcium 
phosphate ($\alpha$-- and $\beta$--TCP) can also be formed during the fabrication of 
CaHAp. Here, we have used a combination of two different methods to produce CaHAp bioceramics, 
which 
were studied by x-ray diffraction (XRD) and infrared (IR) spectroscopy techniques. In addition 
their compression strengths were also measured. The results obtained indicated that 
the mechanical properties depend on the content of CO$_3^{2-}$ ions incorporated into 
the CaHAp lattice, which substitute PO$_4^{3-}$ ions, and possibly on the content of 
Ca$^{2+}$ ions substituted by Mg$^{2+}$ and occasionally Na$^+$ ions.

\section{Experimental Procedures}

\subsection{Preparation of the samples}
\label{secexp}

\subsubsection{Method A}

In the method A the samples (hereafter called A-HAp) were prepared in the following way: 0.090 
mol of (NH$_4$)$_2$HPO$_4$ (Nuclear) and 0.152 mol of Ca(NO$_3$)$_2 \cdot$ 4H$_2$O (Nuclear) were 
completely dissolved in distilled water. Thus, a NH$_4$OH 25vol\% solution is added to the 
previous solution. The obtained solution is then heated at 65$^\circ$C for 90 min under stirring. 
After that, the beaker is sealed, and the solution is heated to its boiling point for 2 h. Then, 
it is cooled to room temperature and the precipitates were allowed to settle overnight. Next, 
they were filtered, washed with distilled water in order to keep a pH 7 and dried at 
60$^\circ$C for 4 h. Some amounts of the A-HAp samples were sintered at 900$^\circ$C in an 
electrically heated box furnace Jung model 0912 for 1 h and they will be called A-HAp-S \cite{Ozgur}.

\subsubsection{Method B}

The samples prepared using method B (hereafter called B-HAp) were produced in the following way: 
first, a solution (hereafter solution B1) was formed by dissolving 0.182 mol of Ca(NO$_3$)$_2 
\cdot $ 4H$_2$O (Nuclear) and 0.0105 mol of MgCl$_2 \cdot $ 6H$_2$O (Reagen) in 500 ml of 
distilled water. Then, a second solution (solution B2) was prepared by dissolving 0.387 mol of 
Na$_2$HPO$_4 \cdot $ 7H$_2$O (Nuclear), 1.25 mol of NaOH (Nuclear), 0.357 mol of NaHCO$_3$ 
(Vetec) and 0.00752 mol of Na$_4$P$_2$O$_7 \cdot $ 10H$_2$O (Resimap) in 1300 ml of distilled 
water. Next, the solutions B1 and B2 were put together in a beaker at room temperature and a 
white precipitate is obtained, which is washed until a pH 8 is reached. Then, the cream was 
dried at 80$^\circ$C for 48 h, and a part of it was also sintered at 900$^\circ$C for 1 h, 
forming the B-HAp-S samples \cite{Lee}.

\subsubsection{Bioceramic Samples}

The bioceramic samples (BIO) were prepared by mixing 1 g of A-HAp, 2 g of B-HAp and 1.8 ml of 
Na$_2$HPO$_4 \cdot$ 7H$_2$O 0.2 M, which acts as an accelerator for the reaction. After that, 
cylindrical samples of 10 mm height and 5 mm diameter were prepared with the obtained cream, and 
they were heat treated at 50$^\circ$C for 2 h. The bioceramic cylinders were also sintered at 
900$^\circ$C for 1 h (BIO-S). In order to perform compression strength comparisons, cylindrical 
samples of A-HAp and B-HAp were made in the same way, and some of them were also sintered at 
900$^\circ$C.

\subsection{X-ray Measurements}

X-ray measurements were collected using a Rigaku power difractometer, Miniflex model, working 
with Cu K$\alpha$ radiation ($\lambda = 1.5418$ \AA). XRD measurements were taken for all samples 
produced. The patterns obtained are shown in Sec. \ref{secres}.

\subsection{Infrared Measurements}

Infrared measurements were performed from 400 to 4000 cm$^{-1}$ using a FTIR Perkin-Elmer 
infrared spectrometer. Spectra were taken for A-HAp, B-HAp, BIO and BIO-S samples.

\subsection{Compression strength measurements}

The compression strength measurements were performed in all samples using a 3M Instron 4444 
equipment.

\section{Results and Discussion}
\label{secres}

\begin{figure}[h]
\begin{center}
\includegraphics{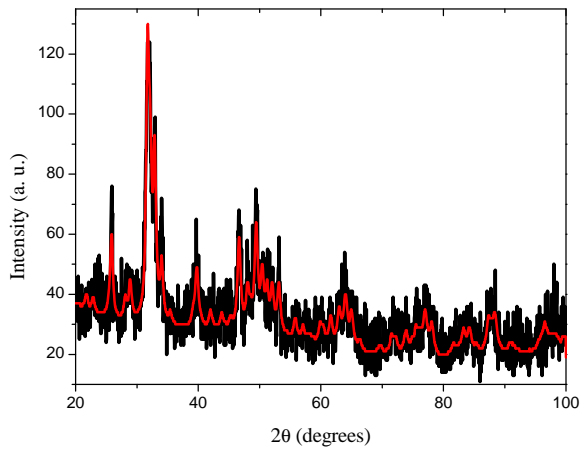}
\caption{\label{fig1} XRD pattern for A-HAp and its Rietveld simulation.}
\end{center}
\end{figure}

\begin{figure}[h]
\begin{center}
\includegraphics{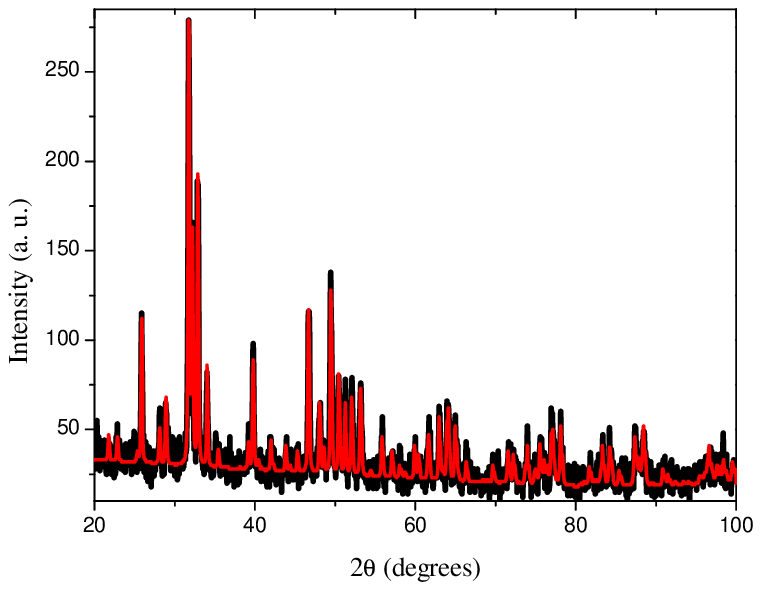}
\caption{\label{fig2} XRD pattern for A-HAp-S and its Rietveld simulation.}
\end{center}
\end{figure}

\begin{figure}
\begin{center}
\includegraphics{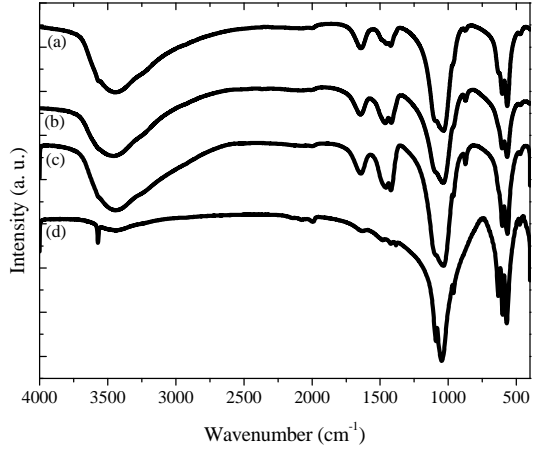}
\caption{\label{fig3} IR spectra for (a) A-HAp; (b) B-HAp; (c) BIO and (d) BIO-S samples.}
\end{center}
\end{figure}

Figure \ref{fig1} shows the x-ray pattern for A-HAp and its simulation using the Rietveld 
refinement procedure \cite{Rietveld}. It was indexed to the hexagonal structure of 
hydroxyapatite given in 
JCPDS card 730294. As it can be seen from this figure, a good agreement was achieved. The 
refined lattice parameters obtained were $a = 9.4207$ \AA\  and $c = 6.8898$ \AA. From the 
refinement and using the Scherrer formula \cite{Scherrer}, 
an average crystallite size of 164 \AA\ was obtained, indicating that it is in the 
nanometric form. No signals of contamination by $\beta$-TCP (tricalcium phosphate) or CaO 
(calcium oxide) were found in this measurement. After the sintering process, its crystallinity 
increases, as it can be seen in figure \ref{fig2}, which shows the x-ray pattern for the 
A-HAp-S sample with its Rietveld refined simulation. Again, there is no contamination by 
$\beta$-TCP or CaO. The new lattice parameters were $a = 9.4236$ \AA\ and $c = 6.8847$ \AA. 
These lattice parameters are different from those given in the JCPDS card 730294 for 
hydroxyapatite ($a = 9.432$ \AA\  and $c = 6.881$ \AA). These differences (decrease in $a$ 
parameter and increase in $c$ parameter) are usually associated with a contamination of the 
CaHAp lattice by CO$_3^{2-}$ ions \cite{Gil}. This contamination could be confirmed by the IR 
measurement performed on the A-HAp sample, shown in fig. \ref{fig3}.a. In this figure, the 
characteristic bands of hydroxyapatite associated with PO$_4^{3-}$ ions \cite{Pleshka} can be seen 
around 470, 567, 603, 1035 and 1100 cm$^{-1}$, and small shoulders can be seen at 967 
cm$^{-1}$, also associated with PO$_4^{3-}$ ions, and at 630 and 3575 cm$^{-1}$, which are 
assigned to OH$^-$ ions \cite{Bayraktar,Pleshka}. The broad bands around 1639 and 3447 cm$^{-1}$ 
are associated with water in the sample \cite{Gil}. The contamination by CO$_3^{2-}$ ions 
gives rises to the bands 
at 870, 1420 and 1465 cm$^{-1}$, and the small bands around 2030 cm$^{-1}$ are associated with 
CO$_2$ vibrational modes, which was probably absorbed from the air \cite{Gil,Morales}. 
The presence of 
CO$_3^{2-}$ ions can be attributed to some contamination of the starting reagents and also to 
the air. Two points should be noted here: first, as biological apatite usually have 
CO$_3^{2-}$ ions, the contamination observed above is not undesirable. Second, in the sintered 
sample much of water and CO$_3^{2-}$ ions are expelled from the sample. The average crystallite 
size of A-HAp-S increases to 536 \AA, making it interesting for several biological applications. 

\begin{figure}[h]
\begin{center}
\includegraphics{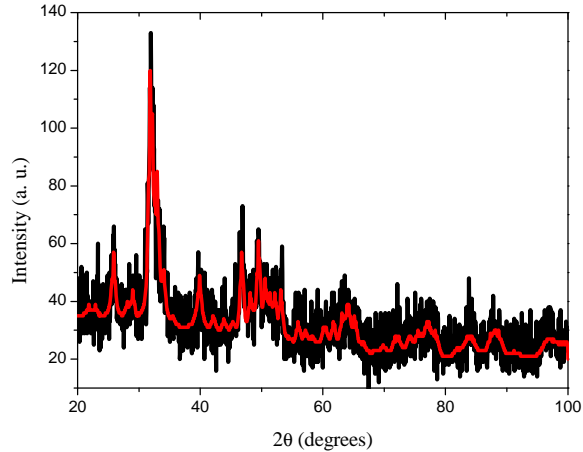}
\caption{\label{fig4} XRD pattern for B-HAp and its Rietveld simulation.}
\end{center}
\end{figure}

Figure \ref{fig4} shows the x-ray pattern for B-HAp. It is very similar to that shown in fig. 
\ref{fig1}, and it was also indexed to the hydroxyapatite phase given in JCPDS card 730294. As 
it happens for A-HAp, no signals of contamination by $\beta$-TCP or CaO were found. Its 
simulation obtained from the Rietveld refinement is also seen in fig. \ref{fig4}, and the 
refined lattice parameters are $a = 9.4023$ \AA\  and $c = 6.8877$ \AA. There is a large 
reduction of the lattice parameter $a$, whereas $c$ increases, in agreement with the fact that 
now the samples were prepared using reagents containing CO$_3^{2-}$ ions, which causes a larger 
substitution of PO$_4^{3-}$ for CO$_3^{2-}$ ions. This substitution produces better-defined bands 
associated with CO$_3^{2-}$ ions, which can be seen in fig. \ref{fig3}.b around 875, 1418 and 
1464 cm$^{-1}$. This figure also shows the bands seen in fig. \ref{fig3}.a for A-HAp. The 
PO$_4^{3-}$ and OH$^-$ bands are found in the same positions, but the water bands change to 
1645 and 3454 cm$^{-1}$, respectively. Again, there is a small contamination by CO$_2$ from air. 
The average crystallite size of this phase is 322 \AA, almost twice of the HAp samples, 
suggesting that the method used in B-HAp preparation is able to produce samples more suitable for 
biological applications than the method used in A-HAp.

\begin{figure}[h]
\begin{center}
\includegraphics{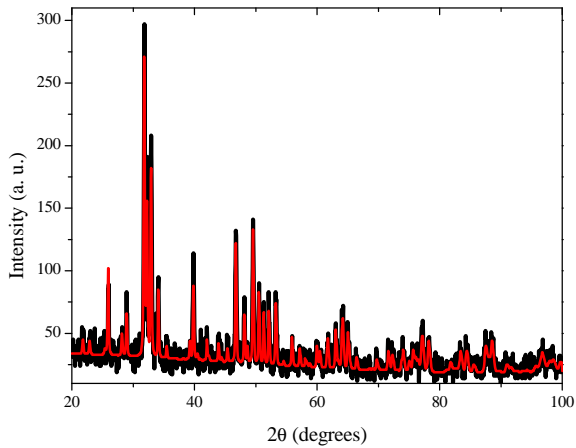}
\caption{\label{fig5} XRD pattern for B-HAp-S and its Rietveld simulation.}
\end{center}
\end{figure}

Figure \ref{fig5} shows the x-ray pattern obtained for B-HAp-S and its simulation obtained using 
the Rietveld refinement procedure. It is similar to the pattern shown in fig. \ref{fig2}, and it 
was also indexed to the hydroxyapatite phase. The refined lattice parameters obtained were 
$a = 9.4213$ \AA\  and $c = 6.8731$ \AA, showing an increase in the lattice parameter $a$ and an 
unexpected large decrease in $c$, which could indicate a larger release of Co$_3^{2-}$ ions by 
this sample. In addition, it is known that Mg$^{2+}$ ions can replace Ca$^{2+}$ ions in the 
hydroxyapatite lattice. Since Mg$^{2+}$ ions are present in the reagents used in method B, this 
fact could be also responsible for these changes in the lattice parameters. The average crystallite 
size of B-HAp-S is 467 \AA, a little smaller than that found for A-HAp-S.

\begin{figure}[h]
\begin{center}
\includegraphics{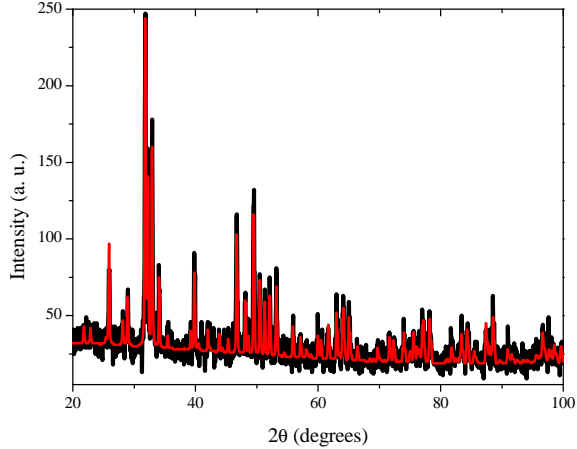}
\caption{\label{fig6} XRD pattern for BIO-S and its Rietveld simulation.}
\end{center}
\end{figure}

Figure \ref{fig6} shows the x-ray pattern measured for the BIO-S samples. It is similar to those 
seen in figs. \ref{fig2} and \ref{fig5}, and corresponds to the hydroxyapatite phase found in 
all x-ray patterns above. Its Rietveld simulation is also shown in fig. \ref{fig6}, and the 
refined lattice parameters are $a = 9.4241$ \AA\  and $c = 6.8825$ \AA. The average crystallite 
size of BIO-S reaches 561 \AA, showing that this is the sample with the highest degree of 
crystallinity. The IR spectra measured for these samples before and after the sintering process 
can be seen in fig. \ref{fig3}.c and \ref{fig3}.d. Fig. \ref{fig3}.c shows the characteristic 
bands of hydroxyapatite (the PO$_4^{3-}$ vibrations at 470, 567, 603, 1036 and 1095 cm$^{-1}$), 
well defined CO$_3^{2-}$ bands around 873, 1420 and 1460 cm$^{-1}$, a OH$^-$ band as a shoulder 
of the very broad band associated to water at 3448 cm$^{-1}$, besides the water band at 1644 
cm$^{-1}$ and the CO$_2$ band at 2000 cm$^{-1}$. After the sintering, the water bands are much 
reduced, and OH$^-$ bands around 633 and 3568 cm$^{-1}$ are much better resolved. The 
CO$_3^{2-}$ bands about 1420 cm$^{-1}$ almost disappeared, indicating a great release of 
CO$_3^{2-}$ ions by the sample.

Although the x-ray measurements of the samples have indicated that the hydroxyapatite phase is 
present in all samples, some differences among them exist, as can be seen by comparing the 
lattice parameters and average crystallite sizes obtained, and also the IR spectra seen in 
fig. \ref{fig3}. However, the most impressive difference is verified in the compression strength 
measurements performed on the samples. 
The cylindrical A-HAp samples have compression strengths ranging from 2 to 5 MPa. After sintering, 
these samples (A-HAp-S) reach compression strengths of 6-9 MPa. The compression strengths of the 
cylindrical B-HAp samples are less than 1 MPa, and do not seem to change after sintering 
(B-HAp-S). Considering now the cylindrical bioceramic samples, their compression strengths changes from 
1-2 MPa before sintering (BIO) to 26--30 MPa after sintering (BIO-S), an increase by almost five 
times when compared to the values found for A-HAp-S and by six times when compared to usual 
hydroxyapatite values reported in the literature \cite{Tadic,Ozgur}. 

A possible explanation for this fact 
is that by mixing the samples produced by the two different methods, the contamination by 
CO$_3^{2-}$, Mg$^{2+}$ and maybe Na$^+$ ions decreased and reached an optimal value, which is 
lesser than that seen in the B-HAp and greater than that found in the A-HAp samples alone. 
Another possibility is that, although B-HAp and B-HAp-S have those contaminants, their porosity 
is too large even after sintering, thus decreasing their compression strength. The increase in 
the compression strength of BIO-S makes it a very good material for biological application in 
dental and bone implants, for instance. A deeper investigation about the causes of this behavior 
is in progress and will be published elsewhere.

\section{Conclusion}

From this study we can conclude that

\begin{enumerate}
\item Calcium hydroxyapatite, Ca$_{10}$(PO$_4$)$_6$(OH)$_2$, was produced by using both methods 
described in Sec. \ref{secexp}, as confirmed by the Rietveld refinements performed on the x-ray 
diffraction patterns of all samples. The infrared spectra agree with this verification and show 
the characteristic bands of hydroxyapatite. The crystallinity of all samples improves after the 
sintering process, as indicated by the increase in the average crystallite sizes.

\item A-HAp samples have a small contamination formed by CO$_3^{2-}$ ions, which is not undesired 
since biological apatite usually shows about 4\% of these ions in its composition. B-HAp 
samples, besides a higher contamination by CO$_3^{2-}$ ions, also have Mg$^{2+}$ and maybe 
Na$^+$ ions in their composition. The contamination by CO$_3^{2-}$ is also confirmed by the 
infrared spectra of the samples, which show the CO$_3^{2-}$ bands.

\item The B-HAp and B-HAp-S samples have a larger porosity than A-HAp and A-HAp-S samples, as 
confirmed by the compression strength measurements, which indicate that B-HAp and B-HAp-S are 
too soft materials. Although the compression strength of A-HAp-S increases after the sintering, 
this value is still low for some dental and bone implant applications.

\item The best samples were obtained by mixing the samples produced by the two methods and 
submitting them to the sintering process, obtaining the BIO-S samples. Their compression 
strengths reach 26--30 MPa, very suitable for dental and bone implant applications. A possible 
explanation for the improvement obtained is that, by mixing the samples produced by the two 
methods, the contamination by CO$_3^{2-}$ and Mg$^{2+}$ ions reached optimal values, increasing 
the mechanical properties of the samples.

\end{enumerate}

\ack

We thank to the Brazilian agencies CNPq and CAPES for financial support.


\begin{thebibliography}{10}
\expandafter\ifx\csname url\endcsname\relax
  \def\url#1{\texttt{#1}}\fi
\expandafter\ifx\csname urlprefix\endcsname\relax\def\urlprefix{URL }\fi

\bibitem{Gil}
G.~Fel\'{\i}cio-Fernandes, M.~C.~M. Laranjeira, Calcium phosphate biomaterials
  from marine algae hydrothermal synthesis and characterisation, Qu\'{\i}mica
  Nova 23 (2000) 441.

\bibitem{Bayraktar}
D.~Bayraktar, A.~C. Tas, Chemical preparation of carbonated calcium
  hydroxyapatite powders at 37$^\circ${C} in urea-containing synthetic body
  fluids, J. Euro. Ceram. Soc. {19} (1999) 2573.

\bibitem{Finisie}
M.~R. Finisie, A.~Josu\'e, V.~T. F\'avere, M.~C.~M. Laranjeira, Synthesis of
  calcium-phosphate and chitosan bioceramics for bone regeneration, An. Acad.
  Bras. Cienc. 73 (2001) 525.

\bibitem{Hench}
L.~L. Hench, J.~Wilson, An Introduction to Bioceramics, World Scientific,
  London, 1993.

\bibitem{Rejda}
B.~V. Rejda, J.~G.~J. Peelen, K.~de~Groot, Tricalcium phosphate as a bone
  substitute, J. Bioeng. {1} (1977) 93.

\bibitem{Jarcho}
M.~Jarcho, Calcium phosphate ceramics as hard tissue prosthetics, Clin. Orthop.
  Relat. Res. {157} (1981) 259.

\bibitem{Rosen}
H.~M. Rosen, Porous, block {HA} as an interpositional bone graft substitute in
  orthognatic surgery, Plast. Reconstr. Surg. {83} (1989) 985.

\bibitem{Passuti}
N.~Passuti, G.~Daculsi, J.~M. Rogez, S.~Martin, J.~V. Bainvel, Macroporous
  calcium phosphate ceramic performance in human spine fusion, Clin. Orthop.
  Relat. Res. {248} (1989) 169.

\bibitem{Byrd}
H.~S. Byrd, P.~C. Hobar, K.~Shewmake, Augmentation of the craniofacial skeleton
  with porous {HA} granules, Plast. Reconstr. Surg. {91} (1993) 15.

\bibitem{Satoh}
K.~Satoh, K.~Nakatsuka, Simplified procedure for aesthetic improvement of
  facial contour by maxillary augmentation using a porous {HA} graft for
  maxillofacial deformity, Plast. Reconstr. Surg. {97} (1996) 338.

\bibitem{Tadic}
D.~Tadic, F.~Peters, M.~Epple, Continuous synthesis of amorphous carbonated
  apatites, Biomaterials {23} (2002) 2553.

\bibitem{Morales}
J.~G. Morales, J.~T. Burgu\'es, T.~Boix, J.~Fraile, R.~R. Clemente,
  Precipitation of stoichiometric hydroxyapatite by a continuous method, Cryst.
  Res. Technol. {36} (2001) 15.

\bibitem{Ozgur}
N.~O. Engin, A.~C. Tas, Preparation of porous
  {C}a$_{10}$({PO}$_4$)$_6$({OH})$_2$ and $\beta$-{C}a$_3$({PO}$_4$)$_2$
  bioceramics, J. Am. Ceram. Soc. {83} (2000) 1581.

\bibitem{Lee}
D. D. Lee, C. Rey, M. Ailova US Pat. 6,214,368 (2001).

\bibitem{Rietveld}
R.~A. Young, A.~Sakthivel, T.~S. Moss, C.~O. Paiva-Santos, {DBWS}-9411 - an
  upgrade of the {DBWS}*.* programs for {R}ietveld refinement with {PC} and
  mainframe computers, J. Appl. Cryst. 28 (1995) 366.

\bibitem{Scherrer}
H.~P. Klug, L.~E. Alexander, X-Ray Diffraction Procedures for Polycrystalline
  and Amorphous Materials, 2nd Edition, John Wiley and Sons, New York, 1974.

\bibitem{Pleshka}
N.~Pleshka, A.~Boskey, R.~Mendelsohn, Novel infrared spectroscopy method on the
  determination of crystallinity of hydroxyapatite minerals, Biophys. J. 60
  (1991) 786.

\end{thebibliography}

\end{document}